\documentclass[aps,prd,showpacs,floatfix,nofootinbib,notitlepage,superscriptaddress,amsthm ,amsmath,amssymb,amsfonts,showkeys]{revtex4-1}

\usepackage{bm}
\usepackage{slashed}

\usepackage{amsmath}    
\usepackage{graphicx}   
\usepackage{subfigure}  
\usepackage{hyperref}   
\usepackage{url}

\usepackage[svgnames]{xcolor}

\usepackage{placeins}
\usepackage{xspace}

\def\LEPT{\texttt{LEPTO}\xspace}
\def\MLEPT{\texttt{\lowercase{m}LEPTO}\xspace}

\def\fS{\varphi_S}

\def\kT{\vect{k}_T}

\def\Pp#1{\vect{P}_{#1\perp}}
\def\PT#1{\vect{P}_{#1T}}

\newcommand{\vect}[1]{\boldsymbol{#1}}

\newcommand{\bfq}{\vect{q}}

\newcommand{\al}[1]{\begin{align} #1 \end{align}}
\newcommand{\non}{\nonumber}
\newcommand{\vf}{\varphi}
\newcommand{\df}{\Delta\varphi}

\newcommand{\Ge}{\mathrm{GeV}}

\begin{document}

\title{Azimuthal correlations in the structure functions of polarized dihadron semi-inclusive deep inelastic scattering}

\author{Aram~Kotzinian}
\affiliation{Yerevan Physics Institute, 2 Alikhanyan Brothers St., 375036 Yerevan, Armenia
}
\affiliation{INFN, Sezione di Torino, 10125 Torino, Italy
}

\begin{abstract}
Azimuthal correlations in two hadron production in deep inelastic scattering of unpolarized leptons off a transversely polarized nucleon are discussed. Specifically, a simple approach for accessing ratios of structure functions corresponding to the Sivers effect and transversity induced single spin azimuthal asymmetries to unpolarized structure functions is presented. This approach is then applied to sample pseudodata generated by a modified version of the LEPTO Monte Carlo event generator that includes the Sivers effect. Using these data, it is shown that the azimuthal correlations in the Sivers-like structure functions and the unpolarized structure functions are significantly different. Collins-like single hadron asymmetries in a dihadron sample are also discussed, for which experimental results were recently presented by the COMPASS collaboration.

\end{abstract}


\date{\today}                                           
\maketitle

\section{Introduction}
\label{SEC_INTRO}

\begin{figure}[h]
\begin{center}
\includegraphics[width=0.5\columnwidth]{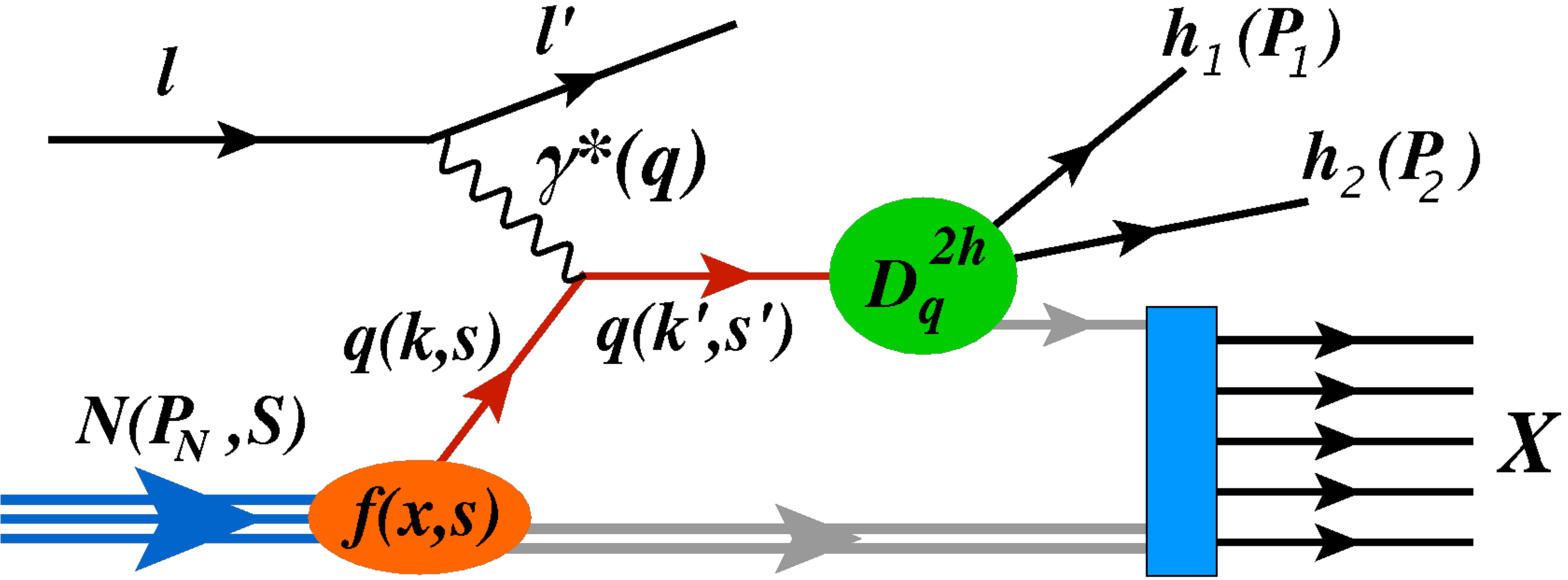}
\caption{The leading order diagram for two hadron production in the current fragmentation region of SIDIS.}
\label{FIG_DIHADRON_CFR}
\end{center}
\end{figure}

The study of correlations between azimuthal angles of two final state hadrons’ momenta in high energy processes plays an important role in understanding the hadronization mechanism. Here, these correlations are considered for the process of two hadron semi-inclusive electroproduction in the current fragmentation region (CFR) of deep inelastic scattering [dihadron semi-inclusive deep inelastic scattering (SIDIS)]
\al{
\label{EQ_2H_SIDIS}
\ell (l) + N({P_N},S) \to \ell (l') + h_1(P_1) + h_2(P_2) + X\,,
}
as schematically depicted in Fig.~\ref{FIG_DIHADRON_CFR}.
%

The expression for the cross-section of this process at the leading twist is long known~\cite{Bianconi:1999cd}. The most general cross-section expression for this process for polarized beam and target, which includes also subleading twist contributions, is presented in a recent work~\cite{Gliske:2014wba}. In both this articles the azimuthal angles of the total and relative momentum of two hadrons were used. However, in experiment the correlations between the azimuthal angles of transverse momenta of each produced hadron were presented, namely as a function of
\al{
\label{EQ_Dphi}
\df=\vf_1-\vf_2,
}
 where $\vf_{1(2)}$ denote the azimuthal angle of the first (second) hadron's transverse momentum (see, for example,~\cite{Arneodo:1986yc}). Very recently, the first preliminary results for the dependence of the Collins-like (CL) asymmetries on $\df$ were presented by the COMPASS collaboration~\cite{COMPASS:2014chia}.

In this paper, we first present the expression for the cross-section of the process (\ref{EQ_2H_SIDIS}) in terms of the transverse momenta (and the corresponding azimuthal angles) of observed hadrons, considering only the unpolarized, Sivers-effect-induced and transversity-induced parts of the cross-section. The one hadron single spin asymmetries (SSAs) in two hadron SIDIS samples are then discussed for both Sivers-like (SL) and CL modulations. Finally, similar to~\cite{Kotzinian:2014lsa,Kotzinian:2014gza}, we use the \LEPT Monte Carlo (MC) event generator~\cite{Ingelman:1996mq}, that has been modified to include the Sivers effect~\cite{Kotzinian:2005zs,Kotzinian:2005zg} to explore the transverse momenta correlations in the kinematical region of the COMPASS experiment in a framework that is independent of the assumptions employed in deriving the analytical expressions for the SSAs.

This paper is organized in the following way. In the next section we briefly introduce the kinematics of the two hadron SIDIS process and present the general expression for the cross-section including the parts induced by the Sivers effect and the combined effect of transversity parton distribution functions (PDFs) and CL dihadron fragmentation functions (DiFFs). Then in Sec.~\ref{SEC_COR_2H} we develop a framework for accessing the $\df$ dependence of the ratios of the polarized to the unpolarized structure functions entering in this expression. In Sec.~\ref{SUB_mLEPTO}, we  briefly discuss the modified LEPTO (\MLEPT) MC generator and apply the proposed framework to the generated pseudodata. In Section~\ref{SEC_Disc} we derive the expressions for the single hadron Collins-like asymmetries in dihadron sample. In Section~\ref{SEC_CONC}, we present our conclusions and outlook. Finally, in the appendix we present the details of the derivations of the cross-section expression. Then it is shown how to express the structure functions through transverse momentum dependent (TMD) PDFs and DiFFs for their arbitrary dependence on transverse momentum.

\section{TWO HADRON PRODUCTION IN THE CFR OF SIDIS}
\label{SEC_2hSIDIS}

In the standard QCD leading order approach used at high energies, we assume that an initial lepton $\ell$ scatters inelastically on a nucleon $N$, where a virtual photon $\gamma^*$ strikes a quark $q$, which then fragments into several hadrons. In the final state, the scattered lepton $\ell'$ and two of the produced hadrons $h_1$ and $h_2$ are detected.
We denote the momenta of the initial and final leptons as $l$ and $l'$, respectively, the momentum of the nucleon $N$ as $P_N$ and the spin $S$, the momenta of the initial and  fragmenting quarks $k$ and $k'$, and the momenta of the detected hadrons as $P_1$ and $P_2$ respectively. We use the standard SIDIS variables:
\al{
\label{EQ_KIN}
&
q=l-l',\,  Q^2 = -q^2, \ x = \frac {Q^2}{2P_N \cdot q},
\\&\non
\, y = \frac {P_N \cdot q}{P_N \cdot l}, \, z_i = \frac{P_N \cdot P_i}{P_N \cdot q}.
}
Here $q$ is the momentum of the virtual photon, $x$ is the usual Bjorken variable for the quark, and $z_i$ is the fraction of the virtual photon's energy in the laboratory system carried by the produced hadron $h_i$.
We adopt the $\gamma^*-N$ center of mass frame, where the $z$ axis is along the direction of the virtual photon momentum $\bfq$. We will define the transverse components of the arbitrary vector $\bf v$  with respect to the $z$ axis with subscript $_T$: ${\bf v}_T$. The $x$ axis is defined along the lepton transverse momenta $\vect{l}_T$ and $\vect{l}'_T$.

In this article we consider only two SSA related to the transverse polarization of the target: one induced by the transverse momentum dependent Sivers effect (SL asymmetries) and the second induced by the combined effect of the transversity PDF and the CL DiFFs of the transversely polarized quark (CL asymmetries). The derivation of the cross-section expression of the process in Eq.~(\ref{EQ_2H_SIDIS}) in terms of the transverse momenta $\PT{1}$, $\PT{2}$ and the corresponding azimuthal angles $\vf_1$, $\vf_2$ of produced hadrons $h_1$ and $h_2$ is presented in Appendix~\ref{SEC_APP}. The fully unintegrated cross section can be written as
\al
{
\label{EQ_2H_XSEC_GEN}
& \frac{d\sigma^{h_1 h_2}}{dx\, d{Q^2}\, d{\fS}\, dz_1\, dz_2\, d^2\PT{1}\, d^2\PT{2}}  =
\sigma_{U}+S_T(\sigma_{S}+\sigma_{C}),
}
where the SL part is given as
\al
{
\label{EQ_2H_XSEC_Siv}
\sigma_{S} = \sigma_{1S}\sin(\vf_1-\fS) + \sigma_{2S}\sin(\vf_2-\fS),
}
and the CL part is
\al
{
\label{EQ_2H_XSEC_Col}
\sigma_{C} = \sigma_{1C}\sin(\vf_1+\fS) + \sigma_{2C}\sin(\vf_2+\fS).
}
Here $S_T$ is the transverse polarization of target $N$, with azimuthal angle  $\fS$.
The structure functions (SFs) $\sigma_{U}, \sigma_{1,2S}$ and $\sigma_{1,2C}$ depend on $x,Q^2,z_1,z_2,P_{1T},P_{2T}$ and $\df$. The explicit expression for these SFs for arbitrary dependence of the PDFs and the DiFFs on the transverse momenta are presented in Appendix~\ref{SEC_APP}.

One can see from the above three equations, that we have two types of azimuthal correlations: the explicit correlations between transverse momenta of the hadrons and the target transverse polarization $\vf_1 \leftrightarrow \vf_S$ and $\vf_2 \leftrightarrow \vf_S$ given by sine modulations (similar to one hadron SIDIS, but for each hadron separately in our case), and also correlations between the transverse momenta of the hadrons $\vf_1 \leftrightarrow \vf_2$ with each other via $\df$-dependence in the unpolarized and the polarized SFs. Note, that since all the SFs depend only on $\cos(\df)$, they are even functions of $\df$.

\section{Accessing $\vf_1 \leftrightarrow \vf_2$ azimuthal correlations in Structure function}
\label{SEC_COR_2H}

Historically, in SIDIS experiments the single spin azimuthal asymmetries were investigated by extracting the amplitudes of sine or cosine modulations from the data. For example, the amplitude of $\sin(\vf_h-\vf_S)$ modulation were extracted for Sivers effect study in single hadron SIDIS, or the amplitude of $\sin(\vf_R+\vf_S)$ modulation in the dihadron transversity study, where where $\vf_R$ is the azimuthal angle of the relative transverse momentum of the two hadrons.

To study the azimuthal correlations between $\vf_1$ and $\vf_2$ in polarized SFs of dihadron SIDIS, one has to keep the cross-section unintegrated over all three azimuthal angles even when integrating over some or all other kinematic variables, as it was recently done for the first time by COMPASS~\cite{COMPASS:2014chia}. Namely, the amplitudes of CL SSA $\sin(\vf_h-\vf_S)$ modulations were presented for positive and negative hadrons in bins of $|\df|$.

In this section the expressions for the single hadron SL and CL asymmetries as functions of $\df$ in the dihadron sample are presented. In what follows, we keep for brevity only the azimuthal angle variables in all the expressions.

\subsection{Correlations in Sivers-like structure functions}
\label{SUB_COR_SIV}

The expression for the SL modulation depending on the first hadron’s azimuthal angle $\df$  is obtained by a simple change of variables $\vf_1,\vf_2 \rightarrow \vf_1,\df$ in Eq.~(\ref{EQ_2H_XSEC_GEN}):
\al
{
\label{EQ_siv1}
\frac{d\sigma^{h_1 h_2}}{d\vf_S d\vf_1 d\df}  =
\sigma_{U}(\df)+S_T\big[\big(\sigma_{1S}(\df) + \sigma_{2S}(\df)\cos(\df)\big)\sin(\vf_1-\fS) - \sigma_{2S}(\df)\sin(\df)\cos(\vf_1-\fS)\big],
}
and, similarly for the SL modulation on the $\vf_2$
\al
{
\label{EQ_siv2}
\frac{d\sigma^{h_1 h_2}}{d\vf_S d\vf_2 d\df}  =
\sigma_{U}(\df)+S_T\big[\big(\sigma_{2S} (\df)+ \sigma_{1S}(\df)\cos(\df)\big)\sin(\vf_2-\fS) + \sigma_{1S}(\df)\sin(\df)\cos(\vf_2-\fS)\big].
}
As one can see from the above equations, for each hadron we have both sine and cosine modulations in Sivers angle $\vf_{1(2)}-\vf_S$ with amplitudes
\al{
\label{EQ_ASiv1sin}
&
A_{1SL}^{\sin(\vf_1-\fS)}(\df) = \frac{\sigma_{1S}(\df) + \sigma_{2S}(\df)\cos(\df)}{\sigma_{U}(\df)},
\\
\label{EQ_ASiv1cos}
&
A_{1SL}^{\cos(\vf_1-\fS)}(\df) = -\frac{\sigma_{2S}(\df)\sin(\df)}{\sigma_{U}(\df)},
\\
\label{EQ_ASiv2sin}
&
A_{2SL}^{\sin(\vf_2-\fS)}(\df) = \frac{\sigma_{2S}(\df) + \sigma_{1S}(\df)\cos(\df)}{\sigma_{U}(\df)},
\\
\label{EQ_ASiv2cos}
&
A_{2SL}^{\cos(\vf_2-\fS)}(\df) = \frac{\sigma_{1S}(\df)\sin(\df)}{\sigma_{U}(\df)}.
}
Note that since the SFs are even functions of $\df$, the above sine asymmetries are even and cosine ones are odd in $\df$ (for $\df \in [-\pi,\pi]$). The asymmetries depend on the ratios of the polarized to the unpolarized SFs,
\al{
\label{EQ_AS1}
&
A_{1S}(\df) = \frac{\sigma_{1S}(\df)}{\sigma_{U}(\df)},
\\
\label{EQ_AS2}
&
A_{2S}(\df) = \frac{\sigma_{2S}(\df)}{\sigma_{U}(\df)},
}
and their measurement gives access to $\vf_1 \leftrightarrow \vf_2$ correlations in the polarized SFs, provided we know the $\df$ dependence of the unpolarized SF.

It is evident that not all asymmetries in Eqs.~(\ref{EQ_ASiv1sin}--\ref{EQ_ASiv2cos}) are independent, since we have only two independent polarized SFs. For example, the second hadron asymmetries can be expressed using the first hadron ones by
\al{
\label{EQ_ASiv1-2sin}
&
A_{2SL}^{\sin(\vf_2-\fS)}(\df) = A_{1SL}^{\sin(\vf_1-\fS)}(\df)\cos(\df) - A_{1SL}^{\cos(\vf_1-\fS)}(\df)\sin(\df),
\\
\label{EQ_ASiv1-2cos}
&
A_{2SL}^{\cos(\vf_2-\fS)}(\df) = A_{1SL}^{\sin(\vf_1-\fS)}(\df)\sin(\df) + A_{1SL}^{\cos(\vf_1-\fS)}(\df)\cos(\df).
}
Thus, by measuring only the first hadron modulations we already have an access to rations of the polarized to the unpolarized SFs
\al{
\label{EQ_AS1-1}
&
A_{1S}(\df) = A_{1SL}^{\sin(\vf_1-\fS)}(\df) + A_{1SL}^{\cos(\vf_1-\fS)}(\df)\cot(\df),
\\
\label{EQ_AS2-2}
&
A_{2S}(\df) = - A_{1SL}^{\cos(\vf_1-\fS)}(\df)\frac{1}{\sin(\df)}.
}
Nevertheless, extracting all the "entangled" asymmetries in Eqs.~(\ref{EQ_ASiv1sin}--\ref{EQ_ASiv2cos}) from data and verifying them against the relations in Eqs.~(\ref{EQ_ASiv1-2sin},\ref{EQ_ASiv1-2cos})  can be useful as a self-consistency test of the experimental analysis.

\subsection{Analysis of Monte Carlo data generated by \MLEPT}
\label{SUB_mLEPTO}

Here we apply the formalism developed in the previous Section to Monte Carlo generated pseudo-data sample. The sample of 2.5 10$^8$ DIS events were generated with a kinematics similar to that of 2h SIDIS at COMPASS experiment~\cite{Adolph:2014fjw} using \MLEPT -- the modified version~\cite{Kotzinian:2005zs}\cite{Kotzinian:2005zg} of the standard DIS even generator \LEPT~\cite{Ingelman:1996mq}. The DIS events were generated with the following cuts: $0.032 < x <0.7,\; 0.1 <y < 0.9,\; Q^2 > 1$ (GeV/c)$^2$ and $W^2> 25$ GeV$^2$. Further, for for each of the oppositely charged hadrons in an event additional kinematic cuts $P_T>0.1~\Ge$ and $z>0.1$ were imposed.
It was demonstrated in~\cite{Kotzinian:2014lsa},~\cite{Kotzinian:2014gza}, that the results obtained with \MLEPT well describe the Sivers single hadron SIDIS asymmetries measured at COMPASS~\cite{Adolph:2012sp}. In this analysis the hadrons in each pair are ordered by charge: the first hadron is the positively charged one.

\begin{figure}[h]
\centering     
\subfigure{\includegraphics[width=0.45\columnwidth]{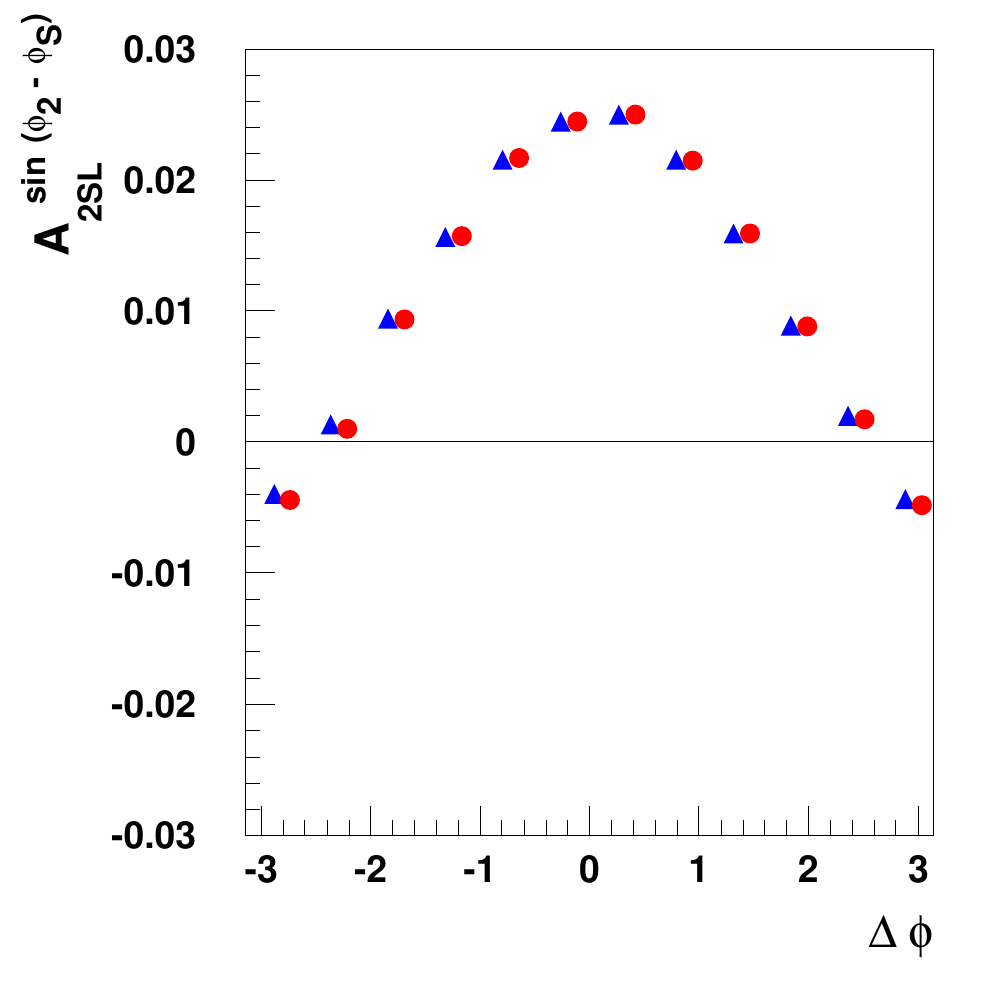}}
\subfigure{\includegraphics[width=0.45\columnwidth]{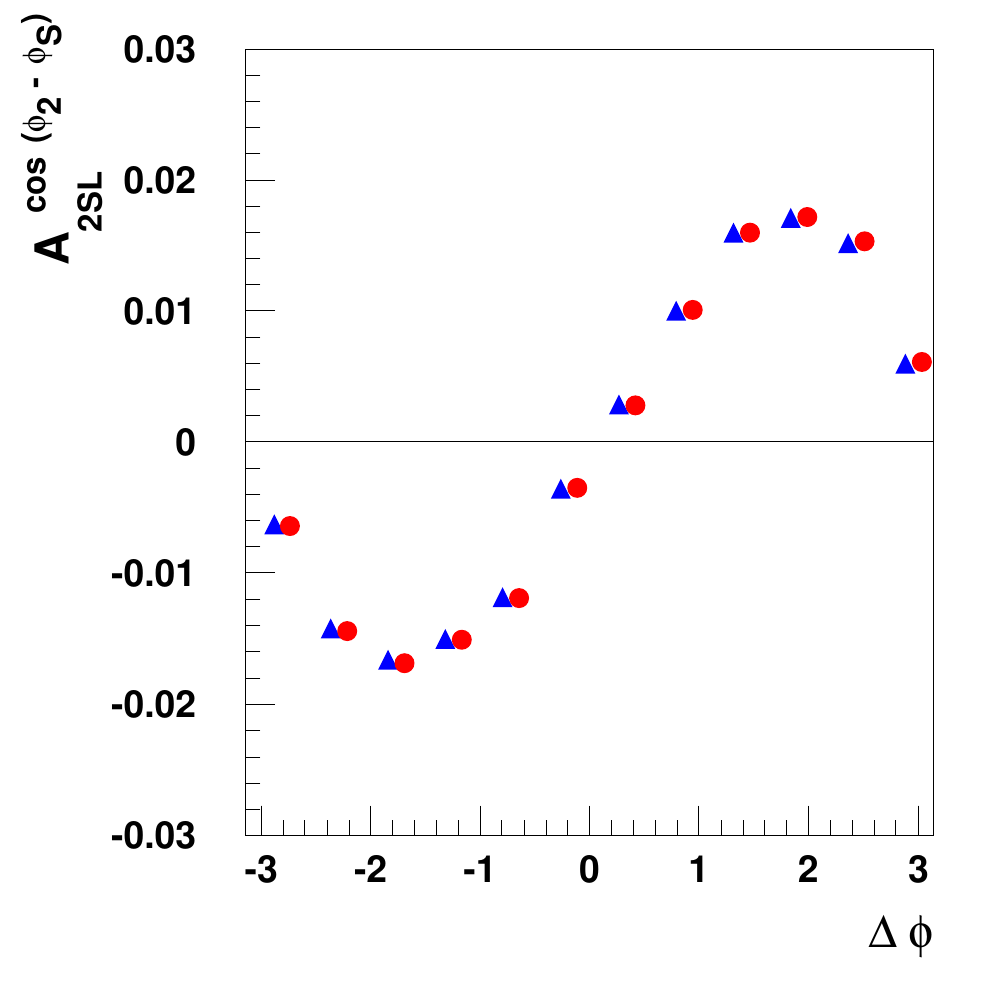}}
\caption{Left: the SL $A_{2SL}^{\sin(\vf_2-\vf_S)}(\df)$ asymmetries (filled triangles) of negative hadron extracted directly from MC data (filled triangles) and calculated  using the entanglement relations from positive hadron sine and cosine asymmetries (filled circles). The calculated asymmetries are plotted shifted on 0.05 rad. Right: the same for $A_{2SL}^{\cos(\vf_2-\vf_S)}(\df)$ asymmetry.}
\label{FIG_Siv2sin-cosFromSiv1}
\end{figure}

In Fig.~\ref{FIG_Siv2sin-cosFromSiv1}, we  present the extracted $A_{1,2SL}^{\sin(\vf_1-\vf_S)}$ and $A_{1,2SL}^{\cos(\vf_1-\vf_S)}$ SL asymmetries using the filled triangle symbols. We also depict using filled circles the corresponding asymmetries obtained form the fits to  the positive hadron SL asymmetries using the entanglement properties of Eqs.~(\ref{EQ_ASiv1-2sin},\ref{EQ_ASiv1-2cos}). It is apparent that the SL asymmetries obtained using both methods agree with each other very well, thus independently validate the relations in the Eqs.~(\ref{EQ_ASiv1-2sin},\ref{EQ_ASiv1-2cos}). In this figures, the errors are smaller than the corresponding marker sizes. Further, in the vicinity of $\df \approx \pm\pi$ they are smaller that at $\df \approx 0$, reflecting the back-to-back correlation of the produced hadrons' transverse momenta in dihadron SIDIS. This feature is well reproduced by \LEPT even generator~\cite{Arneodo:1986yc}.

\begin{figure}[h]
\begin{center}
\includegraphics[width=0.45\columnwidth]{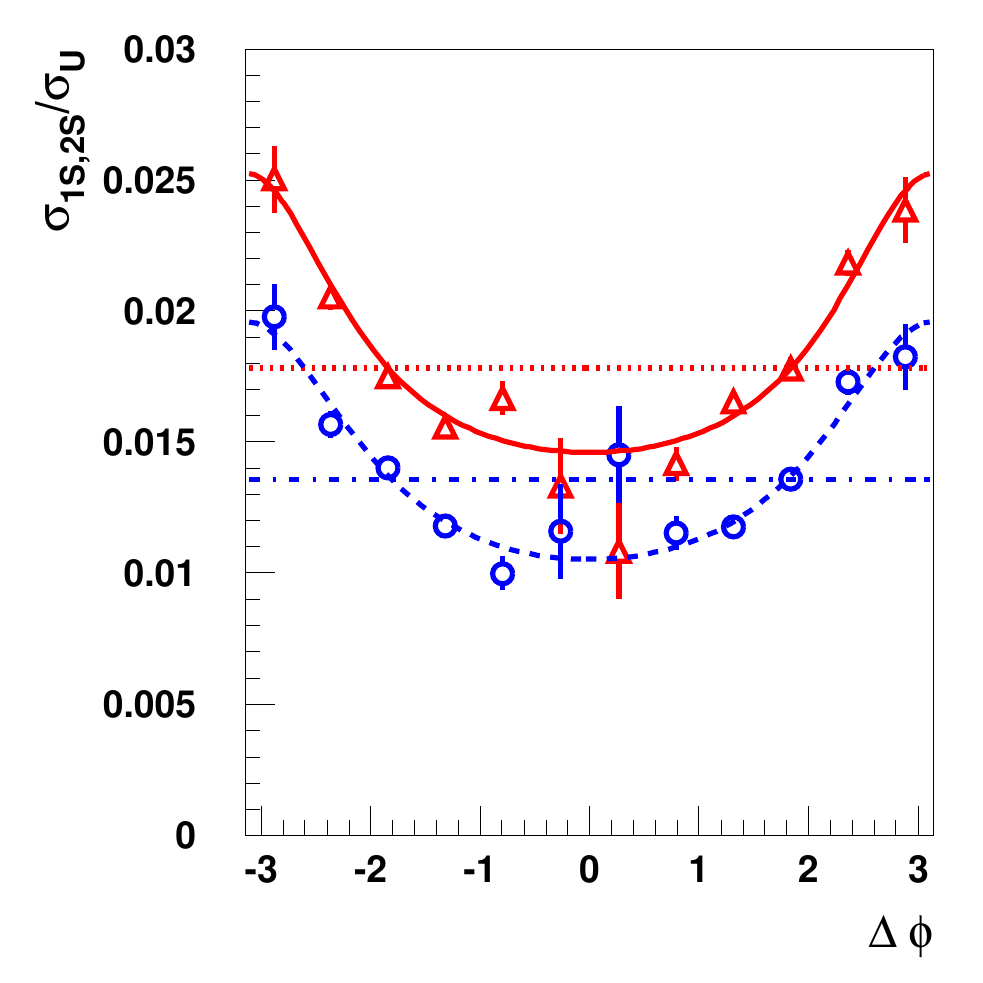}
\caption{The ratios $A_{1S}(\df)$ (triangles) and $A_{2S}(\df)$ (circles) from MC data. The continuous (dashed) line is a result of $\df$-dependent fit and dotted (dot-dashed) lines for constant fit for positive (negative) hadron.}
\label{FIG_Sig12/SigU}
\end{center}
\end{figure}

The ratios of polarized to unpolarized SFs are calculated  using the Eqs.~(\ref{EQ_AS1-1},\ref{EQ_AS2-2}), and then fitted by a function that is either a constant or a ratio of two linear in $\cos(\df)$ functions: $p_1(1+p_2\cos(\df))/(1+p_3\cos(\df))$. Here $p_i$ are the fit parameters. The results are plotted in Fig.~\ref{FIG_Sig12/SigU}. It is clear that the quality of the fits with a $\df$-dependent function is much better than those performed with a constant one.
%

\begin{figure}[h]
\centering     
\subfigure{\includegraphics[width=0.45\columnwidth]{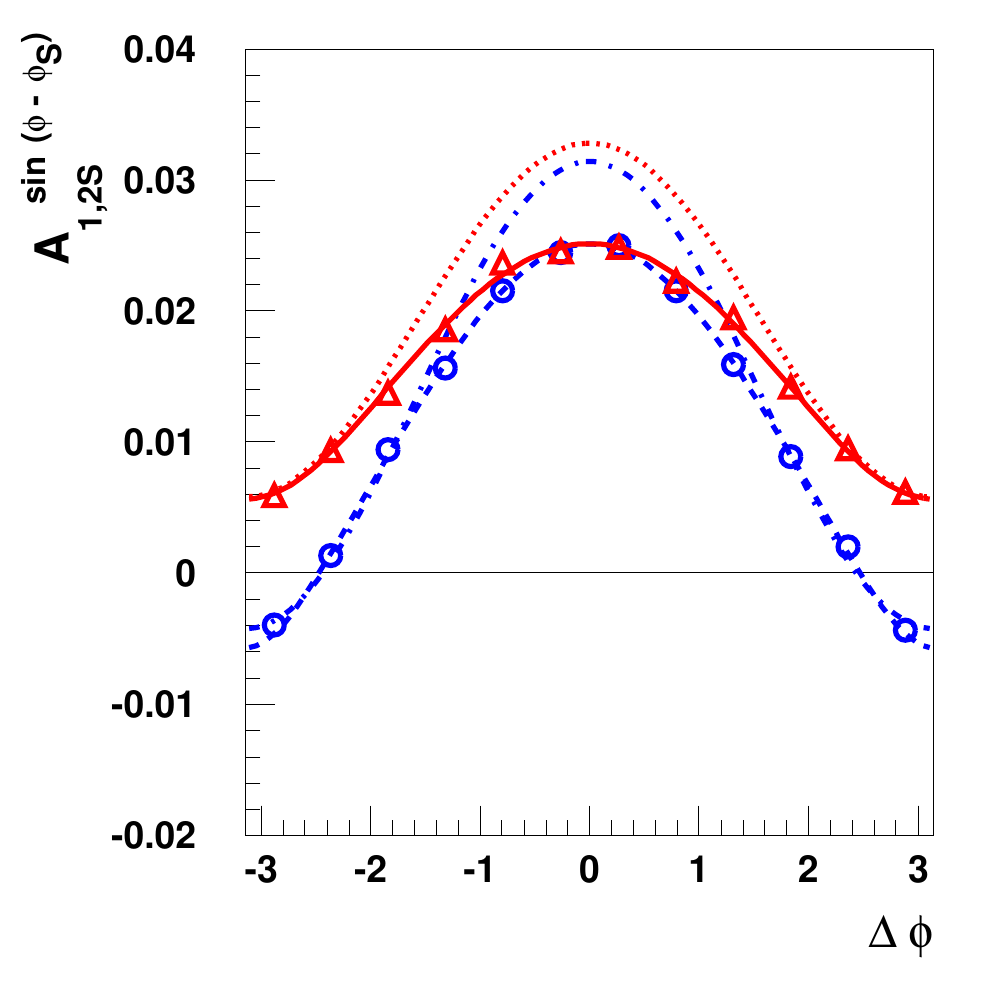}}
\subfigure{\includegraphics[width=0.45\columnwidth]{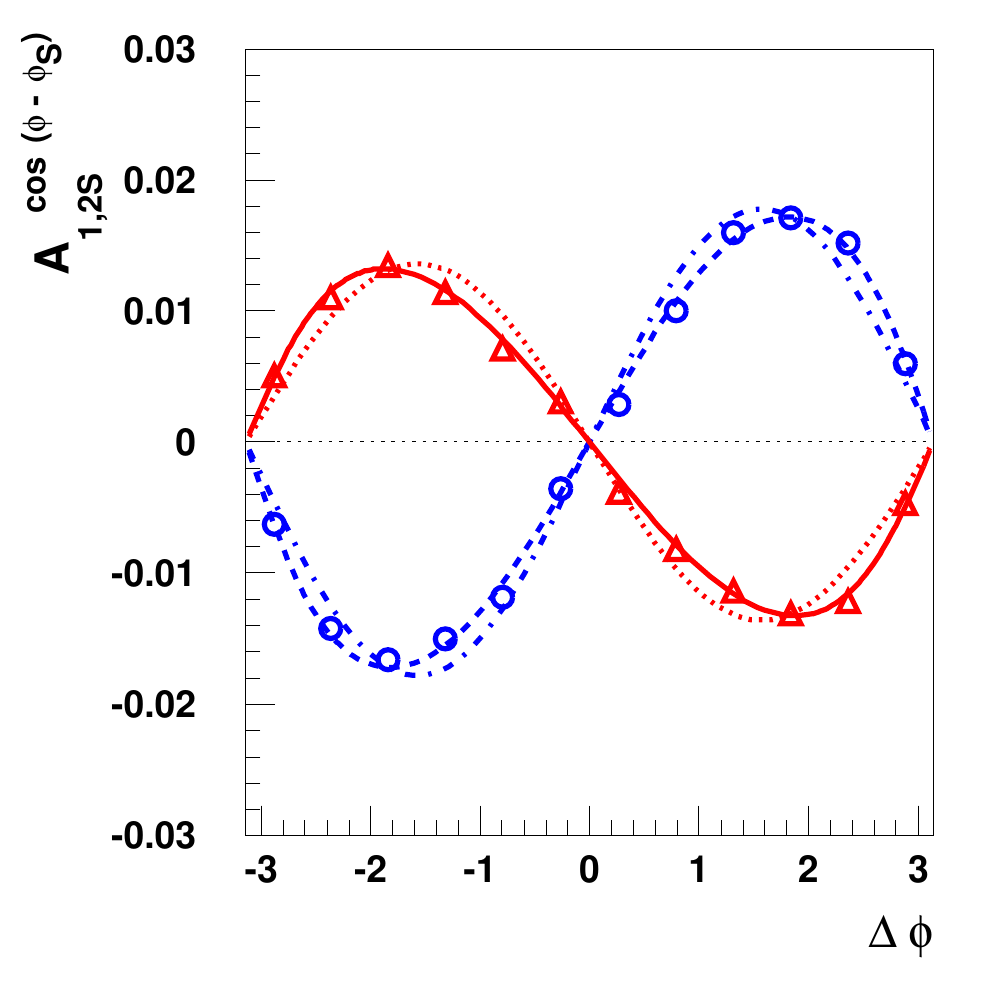}}
\caption{Left: the $A_{1SL}^{\sin(\vf_1-\vf_S)}(\df)$ (triangles) and $A_{2SL}^{\sin(\vf_2-\vf_S)}(\df)$ asymmetries (circles) from MC data. The continuous (dashed) line is a result of $\df$-dependent fit and dashed (dot-dashed) lines for constant fit for positive (negative) hadron. Right: the same for $A_{1SL}^{\cos(\vf_1-\vf_S)}(\df)$ and $A_{2SL}^{\cos(\vf_2-\vf_S)}(\df)$ asymmetries.}
\label{FIG_ASsin-cos}
\end{figure}

In Fig.~\ref{FIG_ASsin-cos} the results for the asymmetries that are directly extracted from the data are shown. The lines correspond to fits with constant and $\df$-dependent functions and using the Eqs.~(\ref{EQ_ASiv1sin} --\ref{EQ_ASiv2cos}). Again, we notice that $\df$-dependent fits better describe the MC data. It should also be noticed, that the explicit calculations using Gaussian parametrization of TMD PDFs and DiFFs show that the $\df$-dependence of Sivers-like and unpolarized SFs should be different, see Appendix in~\cite{Kotzinian:2014gza}.

\subsection{Correlations in Collins-like structure functions}
\label{SUB_COR_COL}

The transversity induced CL sine and cosine asymmetries look very similar to SL ones:
\al{
\label{EQ_ACol1sin}
&
A_{1CL}^{\sin(\vf_1+\fS)}(\df) = \frac{\sigma_{1C}(\df) + \sigma_{2C}(\df)\cos(\df)}{\sigma_{U}(\df)},
\\
\label{EQ_ACol1cos}
&
A_{1CL}^{\cos(\vf_1+\fS)}(\df) = -\frac{\sigma_{2C}(\df)\sin(\df)}{\sigma_{U}(\df)},
\\
\label{EQ_ACol2sin}
&
A_{2CL}^{\sin(\vf_2+\fS)}(\df) = \frac{\sigma_{2C}(\df) + \sigma_{1C}(\df)\cos(\df)}{\sigma_{U}(\df)},
\\
\label{EQ_ACol2cos}
&
A_{2CL}^{\cos(\vf_2+\fS)}(\df) = \frac{\sigma_{1C}(\df)\sin(\df)}{\sigma_{U}(\df)},
}
and all the relations that were presented in Sec.~\ref{SUB_COR_SIV} are also valid in this case (with evident replacements of subscripts $_S \rightarrow  _C$ and $\vf_{1,2}-\fS \rightarrow \vf_{1,2}+\fS$).

Some remark are appropriate here: the preliminary results from COMPASS~\cite{COMPASS:2014chia} were presented only for the sine modulations of the first (positive) and the second (negative) hadrons as functions of $|\df| \in [0,\pi]$~\cite{COMPASS:2014chia}. Using the absolute value of $\df$, we have no access to cosine asymmetries since they are odd in $\df$. Still, one can access the ratios of CL polarized to unpolarized SFs from the extracted sine asymmetries using the following relations
\al{
\label{EQ_R-Col1}
&
\frac{\sigma_{1C}(|\df|)}{\sigma_{U}(|\df|)} = \frac{A_{1CL}^{\sin(\vf_1+\fS)}(|\df|-A_{2CL}^{\sin(\vf_1+\fS)}(|\df|)\cos(|\df|)}{\sin^2(|\df|)},
\\
\label{EQ_R-Col2}
&
\frac{\sigma_{2C}(|\df|)}{\sigma_{U}(|\df|)} = \frac{A_{2CL}^{\sin(\vf_1+\fS)}(|\df|-A_{1CL}^{\sin(\vf_1+\fS)}(|\df|)\cos(|\df|)}{\sin^2(|\df|)}.
}
\FloatBarrier

\section{Single hadron Collins-like asymmetries in the dihadron sample}
\label{SEC_Disc}

In the recently published COMPASS article~\cite{Adolph:2014fjw}, the $x$ dependence of single hadron CL asymmetries in dihadron sample were presented. The remarkable feature of these asymmetries is a mirror symmetry (similarly to single hadron SIDIS) -- the asymmetry for the positive hadron are equal with an opposite sign to the negative hadron one. These asymmetries are integrated over one hadron azimuthal angle, namely over $\vf_2$ in the positive hadron asymmetry and {\it vice versa} for the negative hadron asymmetry. Then the corresponding cross sections are~\footnote{The derivation of this equations is similar to the SL case presented in~\cite{Kotzinian:2014gza}.}

\al{
&\label{EQ_1in2H_1}
\frac{d \sigma^{h_1h_1 } } {dx d\vf_S d\vf_1} = \sigma_{U,0}(x) + S_T \left(\sigma_{1C,0}(x) + \frac{1}{2}\sigma_{2C,1}(x)\right) \sin(\vf_1+\vf_S),
\\&\label{EQ_1in2H_2}
\frac{d \sigma^{h_1h_1 } } {dx d\vf_S d\vf_2} = \sigma_{U,0}(x) + S_T \left(\sigma_{2C,0}(x) + \frac{1}{2}\sigma_{1C,1}(x)\right) \sin(\vf_2+\vf_S),
}
where $\sigma_{U,0}$ and $\sigma_{1(2)C,0(1)}$ denote the moments with respect to $\cos(n\df)$ Fourier expansion of the corresponding SFs:
\al{
&\label{EQ_Fourier}
\sigma(x,\df)_{iC} = \frac{1}{2\pi}\sum_{n=0}^\infty \sigma_{iC,n}(x)\cos(n\df),\quad i=1,2.
}

The mirror symmetry observed in~\cite{Adolph:2014fjw} does not give a direct access to the azimuthal correlations in CL SFs and one cannot conclude that these correlations are similar to that in the unpolarized SF. It only implies
\al{
&\label{EQ_MirrSym}
\sigma_{1C,0}(x) + \frac{1}{2}\sigma_{2C,1}(x) \approx -\sigma_{2C,0}(x) - \frac{1}{2}\sigma_{1C,1}(x),
}
and this approximate equality can be achieved, for example, even if there is no $\df$-dependence in the polarized SFs and $\sigma_{1C,0}(x)=-\sigma_{2C,0}(x)$.

\FloatBarrier
\section{Conclusions}
\label{SEC_CONC}

In this paper we presented for the first time the framework for accessing the azimuthal correlations in polarized structure functions of dihadron SIDIS . The general structure of the cross section in terms of the transverse momentum of each hadron is presented.
It is demonstrated, that a measurement of the sine and the cosine Sivers- and Collins-like asymmetries of one of the hadrons gives access to the ratio of the polarized to the unpolarized structure functions. These asymmetries for the first and the second hadrons are entangled and a simple relations between them is established.

This framework is then applied to a pseudodata sample created by \MLEPT MC generator, which includes the Sivers effect. To validate the asymmetry extraction procedure, the  SL asymmetries for the negatively charged hadron were calculated from those extracted for the positively charged hadrons using the entanglement equations. These results were then compared to those directly extracted from the pseudo-data, verifying that the entanglement properties indeed hold. Then the rations of the polarized SL SFs to the unpolarized SF were calculated from these extracted asymmetries. A better description of this ratio is achieved with a $\df$-dependent function compared to a constant one, which means that the azimuthal correlations in the polarized SL and the unpolarized SFs are different. The origin of this difference can be understood by inspecting the general expressions for the SFs derived in the Appendix: the polarized SFs in Eqs.~(\ref{EQ_2H_SIV_1-2}) contain the additional $\df$-dependent weight factors Eq.~\ref{EQ_2H_weights}) compared to the unpolarized DF Eq.~(\ref{EQ_SIG_U}.

Very recent COMPASS preliminary data~\cite{COMPASS:2014chia} presented only the CL sine asymmetries for positive and negative hadrons in a dihadron sample as a function of $|\df|$. In Sec.~\ref{SUB_COR_COL} it is shown that already with the present COMPASS data we can extract the $\df$-dependence of the rations of the CL polarized to the unpolarized SFs.

In Sec.~\ref{SEC_Disc} it was shown that the recent COMPASS observation~\cite{Adolph:2014fjw} of the mirror symmetry of positive and negative hadron CL asymmetries only establishes a relations between the first two Fourier amplitudes of CL SF and cannot give a direct access to $\df$-dependence of these SFs.

Finally, in the Appendix we give the arguments about the similarity of $\df$-dependence of CL SFs inspired by the observed mirror symmetry~\cite{Adolph:2014fjw} of positive and negative hadrons CL asymmetries.

We conclude that the study of $\df$-dependences of SL and CL asymmetries can play an important role in phenomenological models development since these asymmetries are sensitive to parameters of TMD PDFs and DiFFs.

\section*{Acknowledgements}

I would like to thank my colleagues in the COMPASS collaboration for stimulating discussions and also
Hrayr Matevosyan for careful reading of a draft of this article and valuable comments.

\appendix

\section{General Expression for Structure Functions for arbitrary TMD PDFs and FFs}
\label{SEC_APP}

In this Section we derive explicit expressions for the cross section terms entering in Eqs.~(\ref{EQ_2H_XSEC_GEN}, \ref{EQ_2H_XSEC_Siv}, \ref{EQ_2H_XSEC_Col}).

Let us consider the parts of the cross section for the process in Eq.~(\ref{EQ_2H_SIDIS}) induced by  Sivers and transversity effects. They factorize into sums of convolutions of non-perturbative quark PDFs inside of the transversely polarized nucleon, $f_1^q, f_{1T}^{ \perp q}, h_1^q$,  unpolarized and CL DiFFs for each hadron type, $D_q^{h_1h_2}, H_{1q}^{h_1h_2}, H_{2q}^{h_1h_2}$, and a hard lepton-quark scattering cross-section
\al
{
\label{EQ_2H_XSEC}
\non
\frac{d\sigma^{h_1 h_2}}{dx\, d{Q^2}\, d{\fS}\, dz_1\, dz_2\, d^2\PT{1}\, d^2\PT{2}} & =C(x,Q^2)
\sum_q e_q^2 \int d^2 \kT\  \Bigg[f_{1q}(x, k_T)\ D_q^{h_1h_2}(z_1, z_2, \vect{P}_{1\perp}, \vect{P}_{2\perp})
\\ \non &
+ \epsilon_{i,j}S_T^i\frac{k_T^j}{M}f_{1T}^{\perp q}(x,k_T)\ D_q^{h_1h_2}(z_1, z_2, \vect{P}_{1\perp}, \vect{P}_{2\perp})
\\ & \hspace{-2.5cm}
+ \epsilon_{i,j}s'^j_T h_1^q(x,k_T)\ \left(\frac{P_{1T}^i}{m_1} H_{1q}^{h_1h_2}(z_1, z_2, \vect{P}_{1\perp}, \vect{P}_{2\perp}) + \frac{P_{2T}^i}{m_2} H_{2q}^{h_1h_2}(z_1, z_2, \vect{P}_{1\perp}, \vect{P}_{2\perp})
\right)\Bigg],
}
where $C(x,Q^2) =  \frac{\alpha^2}{Q^4}(1+(1-y)^2)$, $\alpha$ is the fine-structure coupling constant, $M$ is the nucleon mass, $m_{1(2)}$ is the first (second) produced hadron mass, and $\epsilon_{i,j}$ is the two dimensional Levi-Civita tensor.

The first term in the square brackets corresponds to the unpolarized part of the cross section, the second to the Sivers effect induced one and the third one to the transversity and CL fragmentation induced part. Note that in dihadron SIDIS we have two CL spin-dependent DiFF, $H_{1q}^{h_1h_2}$ and $H_{2q}^{h_1h_2}$, {\it cf.}~\cite{Bianconi:1999cd}. Here all TMF PDFs are scalar functions of $x,k_T$ (and $Q^2$), whereas the DiFFs are scalar functions of the energy fractions and the transverse components of the produced hadrons' momenta with respect to the fragmenting quark's momentum (denoted with subscript $_\perp$) $D_q^{h_1h_2}(z_1, z_2, \vect{P}_{1\perp}, \vect{P}_{2\perp})$ (and $Q^2$). In the DIS limit for small transverse momenta, similarly to the one hadron SIDIS case~\cite{Anselmino:2005nn}, the transverse momenta of the hadrons acquired during the fragmentation process are related to the observable $\vect{P}_{1T}, \vect{P}_{2T}$ by
\al
{
\label{EQ_PT}
\Pp{1} \approx \PT{1} - z_1 \kT,\\  \non
\Pp{2} \approx \PT{2} - z_2 \kT.
}
The magnitude and the azimuthal angle of the fragmenting quark's transverse polarization $\bf s'$ in the CL part of Eq.~(\ref{EQ_2H_XSEC}) are~\cite{Kotzinian:1994dv}
\al
{
\label{EQ_s'}
s' = D_{nn}(y)S_T, \quad \vf_{s'}=\pi-\vf_S; \quad D_{nn}(y)=\frac{2(1-y)}{1+(1-y)^2}.
}

To derive a general expression for the cross section for arbitrary transverse momentum dependences of TMD PDFs and DiFF (omitting the dependences on the scalar arguments $x,Q^2,z_1,z_2$ ), it is convenient to introduce the following notation for a convolution of transverse momenta
\al{
\label{EQ_Conv}
{\cal C}\left[g(\vect{k}_T,\vect{P}_{1T},\vect{P}_{2T})\right] = C(x,Q^2)\sum_q e_q^2 {\int {{d^2}{\kT}}g(\vect{k}_T,\vect{P}_{1T},\vect{P}_{2T})},
}
where $g$ can be a scalar or a two-dimensional vector function.

The cross section of Eq.~(\ref{EQ_2H_XSEC}) can be separated into the usual unpolarized part, $\sigma_U$, and a spin-dependent SL $\sigma_S$ and CL $\sigma_C$ parts:
\al{
\label{EQ_SIG}
 \frac{d\sigma^{h_1h_2}}{dx\, d{Q^2}\, d{\fS}\, dz_1\, dz_2\, d^2\PT{1}\, d^2\PT{2}} =  \sigma_U + \sigma_{S}+\sigma_{C}.
}
Using the definition~(\ref{EQ_Conv}), the unpolarized part is
\al{
\label{EQ_SIG_U}
\sigma_U = {\cal C}\left[f_1^q(k_T)\ D_q^{h_1h_2}\left(P_{1\perp},P_{2\perp},\vect{P}_{1\perp}\cdot\vect{P}_{2\perp}\right)\right].
}
For the polarization dependent part of Eq.~(\ref{EQ_2H_XSEC}) the functions $g$ of Eq.~(\ref{EQ_Conv}) are vector functions:
\al{
\label{EQ_g_s-g_c}
g_s^i = \frac{k_T^i}{M} \ f_{1T}^{\perp q}\ D_q^{h_1h_2}, \quad
g_{c1}^i =\frac {P_{1T}^i-z_1k_T^i}{m_1} \ h_1^q\ H_{1q}^{h_1h_2}, \quad
g_{c2}^i =\frac {P_{2T}^i-z_2k_T^i}{m_2} \ h_1^q\ H_{1q}^{h_1h_2}.
}

Let us first consider the SL term. Since ${\cal C}\left[\frac{k_T^i}{M} \ f_{1T}^{\perp q}\ D_q^{h_1h_2}\right]$ can depend only on $\vect{P}_{1T}$ and $\vect{P}_{2T}$ using rotational and parity invariance it is easy to see, that the only possible decomposition looks as
\al{
\label{EQ_g_s_decom}
{\cal C}\left[\frac{k_T^i}{M} \ f_{1T}^{\perp q}\ D_q^{h_1h_2}\right] = \frac{P_{1T}^i}{M}I_{1S}+\frac{P_{2T}^i}{M}I_{2S} .
}
Then we see from Eq.~(\ref{EQ_2H_XSEC}), that the dependence of $\sigma_S$ on the azimuthal correlations $\vf_{1,2} \leftrightarrow \fS$ is given by two SL terms
\al{
\label{EQ_2H_SIV_GEN}
\sigma_{S}
= S_T\big(\sigma_{1S} \sin(\vf_1-\fS) + \sigma_{2S}\sin(\vf_2-\fS)\big),
}
where the SL SFs
\al{
\label{EQ_2H_SIV_1s-2s}
\sigma_{1S}=\frac{P_{1T}}{M}I_{1S}, \quad\sigma_{2S}=\frac{P_{2T}}{M} I_{2S},
}
depend on $x,Q^2,z_1,z_2,P_{1T},P_{2T}$ and ${P}_{1T}{P}_{2T}\cos(\df)$.

In Ref.~\cite{Kotzinian:2014gza}, the explicit calculations of the structure functions $\sigma_{1S}$ and $\sigma_{2S}$ were presented for a Gaussian ansatz for the transverse momentum dependences of PDFs and DiFF. Using Eq.~(\ref{EQ_g_s_decom}), one can find the general expression for SL SFs for arbitrary dependence of PDF and DiFF on transverse momenta:
\al{
\label{EQ_2H_SIV_1-2}
& \sigma_{1S}=\frac{P_{1T}}{M}{\cal C}\left[w_{1}(\vect{k}_T,\vect{P}_{1T},\vect{P}_{2T})\, f_{1T}^{\perp q}(k_T)\, D_{1q}^{h_1 h_2}\left(P_{1\perp},P_{2\perp},\vect{P}_{1\perp}\cdot\vect{P}_{2\perp}\right)\right],
\\ \non
& \sigma_{2S}=\frac{P_{2T}}{M}{\cal C}\left[w_{2}(\vect{k}_T,\vect{P}_{1T},\vect{P}_{2T})\, f_{1T}^{\perp q}(k_T)\, D_{1q}^{h_1 h_2}\left(P_{1\perp},P_{2\perp},\vect{P}_{1\perp}\cdot\vect{P}_{2\perp}\right)\right],
}
where $\vect{P}_{1\perp}$ and $\vect{P}_{2\perp}$ are given by Eq.~(\ref{EQ_PT}) and weights entering in these convolutions are
\al{
\label{EQ_2H_weights}
& w_{1}=\frac{P_{2T}^2\,\left(\vect{P}_{1T}\cdot\vect{k}_{T}\right)-
\left(\vect{P}_{1T}\cdot\vect{P}_{2T}\right)\left(\vect{P}_{2T}\cdot\vect{k}_{T}\right)}
{P_{1T}^2P_{2T}^2-\left(\vect{P}_{1T}\cdot\vect{P}_{2T}\right)^2},
\\ \non
& w_{2}=\frac{P_{1T}^2\,\left(\vect{P}_{2T}\cdot\vect{k}_{T}\right)-
\left(\vect{P}_{1T}\cdot\vect{P}_{2T}\right)\left(\vect{P}_{1T}\cdot\vect{k}_{T}\right)}
{P_{1T}^2P_{2T}^2-\left(\vect{P}_{1T}\cdot\vect{P}_{2T}\right)^2}.
}
As one can readily see from Eqs.~(\ref{EQ_SIG_U},\ref{EQ_2H_SIV_1-2},\ref{EQ_2H_weights}), the convolutions describing unpolarized SF and SL SFs are different -- SL structure functions contain additional weights which are absent in unpolarized SF. We also see from~(\ref{EQ_2H_weights}) that these weights explicitly depend on $\df$ and also on relative azimuthal angles of one of the hadrons and quark transverse momenta. Thus, it is natural to expect that $\df$-dependence in SL SFs and unpolarized SF should be different. This conclusion is also confirmed by the explicit calculations in Gaussian model for TMD PDF and FFs considered in~\cite{Kotzinian:2014lsa},~\cite{Kotzinian:2014gza}. Moreover, in this model even in the case when there is no azimuthal correlation between transverse momenta in DiFF, $\Pp{1}$ and $\Pp{2}$, the azimuthal correlations between $\PT{1}$ and $\PT{2}$ after $\vect{k}_{T}$-integration appears to be different in the SL and unpolarized SFs due to difference of transverse momentum widths in SL and unpolarized PDFs.

Similarly, repeating the derivation analogous to those for SL SFs, one one obtains the following expression for the transversity induced part of cross-section for the process Eq.~(\ref{EQ_2H_SIDIS}
\al{
\label{EQ_2H_COL_GEN}
\sigma_{C}
= S_T\left(\sigma_{1C} \sin(\vf_1+\fS) + \sigma_{2C} \sin(\vf_2+\fS)\right),
}
where the CL SFs are
\al{
\label{EQ_2H_COL_1}
& \sigma_{1C}=D_{nn}(y)\frac{P_{1T}}{m_1} \left( {\cal C}\left[h_1^q H_{1q}^{h_1h_2}\right] -
{\cal C}\left[w_{1}h_1^q \left(z_1H_{1q}^{h_1h_2}+\frac{m_1}{m_2}z_2H_{2q}^{h_1h_2}\right)\right]\right),
\\ \label{EQ_2H_COL_2}
& \sigma_{2C}=D_{nn}(y)\frac{P_{2T}}{m_2}\left( {\cal C}\left[h_1^q H_{2q}^{h_1h_2}\right] -
{\cal C}\left[w_{2}h_1^q \left(z_2H_{2q}^{h_1h_2}+\frac{m_2}{m_1}z_1H_{1q}^{h_1h_2}\right)\right]\right).
}
We see that SF $\sigma_{1C}$ is given by the sum of two parts: the first term is similar to unpolarized DiFF and contains only transversity TMD PDF $h_1^q$ and $H_{1q}^{h_1h_2}$ and the second term contains the weight functions $w_1$ and contribution from both DiFFs $H_{1q}^{h_1h_2}$ and $H_{2q}^{h_1h_2}$, and similarly for $\sigma_{2C}$ substituting $1 \leftrightarrow 2$. Since the weights $w_{1,2} \propto {\vect k}_T$ these terms would vanish if the transversity PDF would be collinear: $h_1^q(x,k_T) \propto \delta^2({\vect k}_T)$.

For the case of oppositely charged pion pair production we can give qualitative arguments in the favour of smallness of the second terms in the Eqs.~(\ref{EQ_2H_COL_1},\ref{EQ_2H_COL_2}), when using a more realistic choice for the transversity PDFs. First of all, the mirror symmetry was observed in single hadron SIDIS Collins asymmetries at HERMES~\cite{Airapetian:2010ds} and COMPASS~\cite{Adolph:2012sn} and also in single hadron asymmetries in dihadron SIDIS~\cite{Adolph:2014fjw}. The favored and unfavored Collins FFs extracted from combined analysis of asymmetries in SIDIS and in two back-to-back hadron pair production in $e^+e^-$ The favored and unfavored Collins FFs extracted from combined analysis of asymmetries in SIDIS and in two back-to-back hadron pair production in~\cite{Anselmino:2007fs}. Finally, in the recent work~\cite{Matevosyan:2013eia}, where the NJL-jet model was used to calculate the Collins FFs, the mirror symmetry was also established between the positive and negative pion Collins FFs. Now, assuming a "mirror symmetry" of CL DiFFs, $H_{1q}^{h_1h_2} \approx -H_{2q}^{h_1h_2}$, and also that for $z$-integrated cross section of oppositely charged pion pair production $\langle z_1 \rangle \approx \langle z_2 \rangle$ (as it is the case for our pseudo-data sample), we see that the convolutions containing the weighs are vanishing. The remaining convolutions are identical to that of unpolarized SF, suggesting that the $\df$-dependence of the $z$-integrated CL SFs can be closer to unpolarized SF one than in the SL SFs case.

\FloatBarrier

\bibliographystyle{apsrev}
\bibliography{fragment}

\end{document}